\documentclass{article}
\usepackage{amssymb,amsfonts,amsmath}
\usepackage[dvips]{graphicx}
\usepackage[hmargin=3cm,vmargin=3.5cm]{geometry}
\usepackage{authblk}

\begin{document}

\title{Emergence of the advancing neuromechanical phase in a resistive force dominated medium}

\author{Yang Ding$^{1,*}$, Sarah S. Sharpe$^2$, Kurt Wiesenfeld$^1$ and
Daniel I. Goldman$^{1,2,**}$ \\
1. School of Physics,    \\
2. Interdisciplinary Bioengineering Program, \\
Georgia Institute of Technology, Atlanta, United States \\
* (current) Department of Aerospace and Mechanical Engineering, \\ University of Southern California, Los Angeles, USA \\
** To whom correspondence should be addressed. \\
 837 State Street NW, Atlanta, GA 30332-0430, \\
 E-mail: daniel.goldman@physics.gatech.edu }

\maketitle

\renewcommand{\thefootnote}{\fnsymbol{footnote}}


\begin{abstract}
Undulatory locomotion, a gait in which thrust is produced in the opposite direction of a traveling wave of body bending, is a common mode of propulsion used by animals in fluids, on land, and even within sand. As such it has been an excellent system for discovery of neuromechanical principles of movement. In nearly all animals studied, the wave of muscle activation progresses faster than the wave of body bending, leading to an advancing phase of activation relative to the curvature towards the tail. This is referred to as ``neuromechanical phase lags'' (NPL). Several multi-parameter neuromechanical models have reproduced this phenomenon, but due to model complexity the origin of the NPL has proved difficult to identify. Here we use perhaps the simplest model of undulatory swimming to accurately predict the NPL during sand-swimming by the sandfish lizard, with no fitting parameters. The sinusoidal wave used in sandfish locomotion, the friction-dominated and non-inertial granular resistive force environment, and the simplicity of the model allow detailed analysis, and reveal the fundamental mechanism responsible for the phenomenon: the combination of synchronized torques from distant points on the body and local traveling torques. This general mechanism should help explain the NPL in organisms in other environments; we therefore propose that sand-swimming could be an excellent system to quantitatively generate and test other neuromechanical models of movement. Such a system can also provide guidance for the design and control of robotic undulatory locomotors in complex environments.
\end{abstract}

%







Animal movement emerges from the complex interplay of nervous and musculoskeletal systems with the environment. Much progress has been made for understanding the neural control patterns and motor systems responsible for effective locomotion (\cite{holAful06,ijspeert20082008,alexanderbook,nishikawa2007neuromechanics,orlovsky1999neuronal,josephson1985mechanical,hof1984emg}). While the environment's influence on neural control is increasingly recognized~\cite{tytell2011spikes,chiel2009brain}, challenges remain in understanding how environments shape the control strategy of locomotion. Particular behaviors, gaits and environments have revealed themselves to be amenable to detailed comparison of experiment and theory to elucidate neuromechanical principles of control \cite{nishikawa2007neuromechanics,tytell2010interactions,dicAfar,LABid2660138}. A form of locomotion where there has been much progress is {\em undulatory locomotion}, a movement strategy employed by numerous, phylogenetically diverse animals such as fish, snakes, worms and sandfish lizards (\textit{Scincus scincus}) (see Fig. 1)~\cite{cohen2010swimming,gray1955propulsion,hu2009mechanics,jayne1995speed,sfakiotakis1999review, maladen2009undulatory} to traverse fluids, solids, and even sand.

In undulatory locomotion, a traveling wave of muscle activation (and curvature) propagates from head to tail resulting in forward movement. The forces produced on different ``segments'' of the body can be decomposed into thrust and drag, and integrating these over the body at any instant in time determines the propulsion of the animal. Many robots have also been built that use such a gait~\cite{wright2007design,crespi2008online,choset2000design}. A feature of undulatory locomotion that is observed across a range of animal sizes and environments is that the wave of muscle activation travels faster than the wave of curvature ~\cite{jayne1995red,wardle1995tuning,gillis1998environmental,williams1989locomotion,sharpe12}. Consequently, the relative phase of the muscle activation to the curvature advances along the body. Physically, this means that more posterior muscles begin activating earlier in the muscle strain cycle (i.e. while the muscle is lengthening) and produce more negative work than anterior muscles. The phenomenon of the advancing neuromechanical phase is often referred to as the ``neuromechanical phase lags'' (Fig. 1c\,\&\,d), or ``NPL'' for short.

Two complementary modeling approaches are used to understand movement principles. The ``bottom-up'' approach (referred to as ``anchoring'' in \cite{fulAkod}) integrates realistic models of multiple bio-components and the complex interactions among them, as well as the with models of the environment. For example, a model (see ~\cite{mcmillen2008nonlinear}) might incorporate tens to hundreds of muscles, hundreds to thousands of neurons, chemical kinetics, and the nonlinear couplings among them. Further complexity could be added by coupling these models to fluids which are governed by complex partial differential equations. In contrast, the ``top-down'' approach (referred to as ``templates'' in \cite{fulAkod}) identifies coordinated components as one single element to generate reduced models and seeks general principles of system behavior. Using the first approach, many multi-parameter neuromechanical models~\cite{cheng1998continuous,bowtell1991anguilliform,tytell2010interactions,chen2011mechanisms,mcmillen2008nonlinear,pedley1999large,fang2010biomechanical} have been proposed to model undulatory locomotion. While such models qualitatively reproduce the NPL in undulatory swimming, due to uncertainties about the passive body properties and the hydrodynamical forces, as well as the model complexity and number of parameters, it remains a challenge to explain the origin of the phenomenon.

\begin{figure}
\centering
\includegraphics[width=0.7\textwidth]{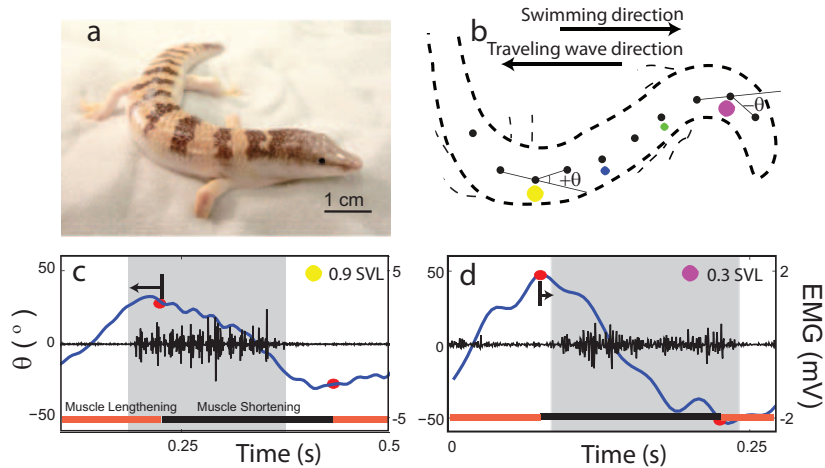}
\caption{Neuromechanical phase lags of the sandfish during sand-swimming. (a) A sandfish lizard resting on 0.3\,mm diameter glass particles. (b) A trace of an x-ray image of the sandfish during subsurface sand-swimming at time, t = 0.21\,s in panels c and d. Opaque markers (black circles) are attached to the exterior midline to facilitate tracking. Electrodes are implanted in epaxial musculature on the right side of the body at 0.3 (magenta), 0.5 (green), 0.7 (blue), 0.9 (yellow) snout-vent length (SVL) locations (where the vent is just posterior to the pelvic girdle and the SVL is approximately 0.75 of the total body length). (c) \& (d) EMG recordings at 0.9 and 0.3 SVL, respectively, during sand-swimming. Gray regions indicate time duration over which the rectified filter EMG is above a threshold (equal to the mean of the rectified-filtered signal) indicating muscle activation (see ~\cite{sharpe12} for more details). The blue line shows the measured angle between consecutive markers (see panel b). The red circles show the maximum or minimum of the best 2nd order polynomial  fit to the angle vs. time series for each half cycle. Arrows indicate the difference in time between the onset of muscle activation and maximal convexity. Note the different scales for EMG due to different electrode constructions. \label{fig:intro}}
\end{figure}

In this paper we show that what might seem to be a specialized and complex system, a lizard ``swimming'' in sand using an undulatory gait, facilitates {\em quantitative} comparison of experiment and theory, and helps explain the fundamental origin of the NPL in undulatory locomotion in other environments. We base the present work on our previous biological muscle activity measurements~\cite{sharpe12} which revealed that the sandfish displays neuromechanical phase lags when targeting a particular behavior--escape. Using a template approach--inputting kinematics of the lizard which confer swimming speed and energetic benefits~\cite{ding2012mechanics} into a previously developed granular resistive force model of sand-swimming and abstracting the nervous system and musculoskeletal system as a ``black box''-- we are able to reproduce internal torque timing patterns (i.e. from muscle contractions) with no fitting parameters. The simple kinematics combined with the relatively simple rheological features of organism-fluidized sand allow us to analyze the model and thus make statements about general principles of neuromechanics in swimming, applicable to organisms and robots in other environments.


\section{Model}
\begin{figure}
\centering
\includegraphics[width=0.8\textwidth]{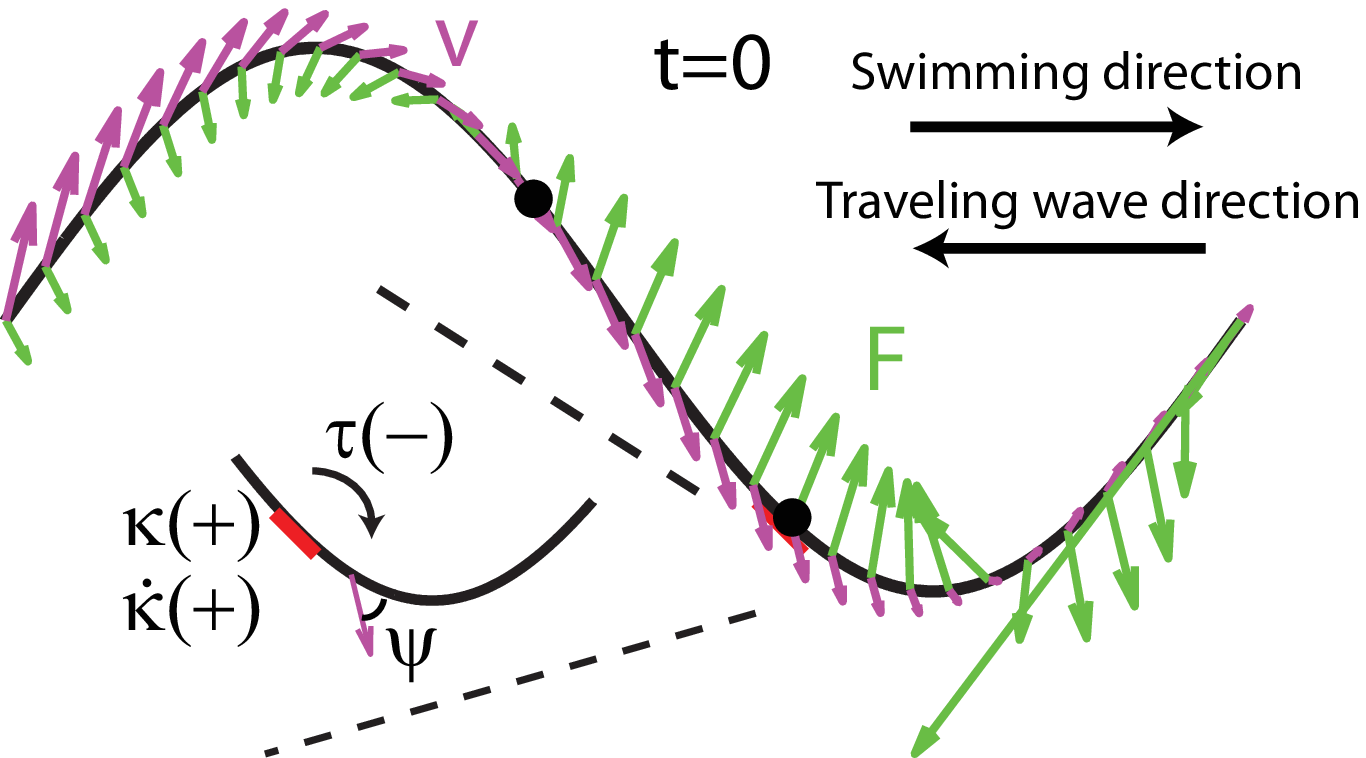}
\caption{Diagram of the model. Magenta arrows represent velocity and green arrows represent forces from the medium. Inset shows the signs of the torque ($\tau$), the curvature ($\kappa$), and the rate of change of curvature ($\dot{\kappa}$) at approximately 0.6 body length on the body. Negative $\tau$  corresponds to no muscle activation on the right side of the body (red thick line). $\psi$ indicates the angle between the segment axis and its velocity. \label{fig:model_diagram}}
\end{figure}
\subsection{Resistive force theory model}
Previously we developed a granular resistive force theory (RFT) model and a numerical simulation that explained the swimming performance of the sandfish~\cite{maladen2009undulatory,maladen2011mechanical,ding2012mechanics}. The models showed that the lizards swim within a self-generated ``frictional fluid'' where frictional forces between the granular particles dominate over both the body inertia and inertial forces from the environment. As before, we prescribe body kinematics (in the frame of the animal) based on the experimental observation that the body position of a sandfish in the body frame is approximately a single-period sinusoidal wave traveling posteriorly (Fig.~\ref{fig:model_diagram}):
\begin{equation}
y_b = A\sin [ 2\pi(\frac{x_b}{\lambda}+\frac{t}{T})], \\
\label{travelingwave}
\end{equation}

\noindent where $y_b$ is the lateral displacement from the midline of a straight animal, $A$ is the amplitude, $T$ is the period of undulation, $\lambda$ is the wavelength, $t$ is the time, and $x_b$ is the distance along a line parallel to the direction of the traveling wave measured from the tail tip. Here, we normalize both the wavelength and period to $2\pi$ such that $\frac{2\pi}{\lambda}=\frac{2\pi}{T}=1$. Since the trunk of the sandfish is quite uniform (with body width variations less than about 5\% from 0.1 snout to vent length (SVL) to 1.0 SVL) and the diameter of the body decreases significantly after about 1.2 SVL, we used a uniform body shape and took the total arc length ($L$) in the model to be 1.2 times the average SVL (8.9 $\pm$ 0.3\, cm) of the animal. Dissection revealed after 1.2 SVL the tail is composed of mostly adipose tissue and a small amount of muscle; therefore, both the external and internal torques on the tail should be minimal for the tail beyond 1.2 SVL. We neglected the variation of the horizontal position $x_b$ of a segment within a cycle, so the normalized position on the animal body $s\frac{2\pi}{L}$ corresponded to the horizontal position $x_b$ in the model, where $s$ is the arc length from the tail end. When a smaller amplitude was used, the wavelength was kept as $2\pi$.

For swimming in sand, the granular force $\vec{F}$ on any infinitesimal segment of the swimmer is independent of the segment speed (and thus undulation frequency), proportional to its depth, and is a function of the angle ($\psi$ in Fig.~\ref{fig:model_diagram}) between the segment axis and its velocity direction.  See Figure~S1 and Equation~S1 in SI for the empirically determined granular force $\vec{F}(\psi)$. The depth of a segment is calculated assuming the model sandfish swims with its center 3.5\,cm below the horizontal plane and at an entry angle of 22 degrees (an average value for the sandfish \cite{maladen2009undulatory}). The entry angle is the angle between the horizontal plane and the plane in which the animal moves~\cite{sharpe12}.


Since the estimated inertial force is negligible, the swimmer moves in a way such that net external force and torque are approximately zero. In this study, we consider all three degrees of freedom in a plane, namely the forward (the only degree of freedom in our previous RFT models), lateral and yaw motion (``recoil''), and determine the velocities of the three degrees of freedom by solving the force/torque balance equations at every instant of time. Since the motion of a point on the body is the superposition of the prescribed and center of mass (CoM) motions, the net external force $\vec{F}_\mathrm{net}(\dot{\vec{R}},\dot{\theta})$ and net external torque about the CoM $\vec{\tau}_\mathrm{net}(\dot{\vec{R}},\dot{\theta})=(0,0,\tau_{\mathrm{net}})$  are functions of the center of mass velocity $\dot{\vec{R}}$ and rotation rate about the CoM $\dot{\theta}$.
For the CoM movement, Newton's laws give
\begin{equation}
\begin{aligned}
\vec{F}_\mathrm{net}(\dot{\vec{R}},\dot{\theta})&=M \ddot{\vec{R}} \\
\vec{\tau}_\mathrm{net}(\dot{\vec{R}},\dot{\theta})&=\dot{\vec{L}}
\end{aligned}
\label{newton}\end{equation}
\noindent where $M$ is total mass and $L$ is angular momentum. By setting the inertial terms on the right sides of these equations to zero, the center of mass velocities ($\dot{\vec{R}}$ and $\dot{\theta}$) can be numerically determined.

\begin{figure}
\centering
\includegraphics[width=0.6\textwidth]{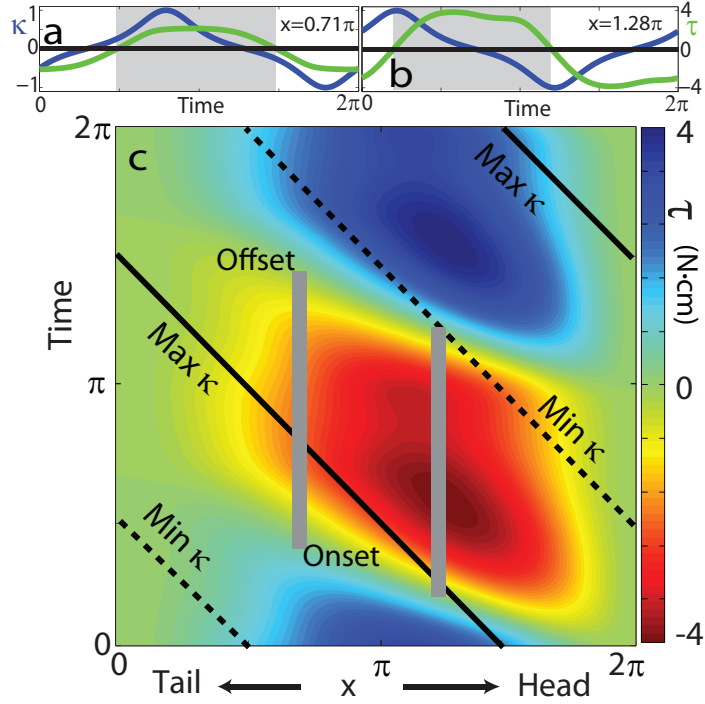}
\caption{Neuromechanical phase lags in the model. (a) \& (b) The curvature (blue lines), torque (green lines), and the predicted muscle activation (gray shaded region) from the RFT model at two representative points indicated by black dots in Figure~\ref{fig:model_diagram} and gray vertical lines in (c). (c) Torque as a function of time and position along the body. Gray vertical bars indicate the predicted muscle activation durations at two representative points. Solid and dashed black lines represent the time when the maximal curvature and minimal curvature are reached, respectively. \label{fig:model}}
\end{figure}
\subsection{Torque calculation in RFT}
Because inertia is negligible, the net torque due to the granular force on a portion (e.g. $[x_b,\,2\pi]$) of the sandfish body about any point of interest $x_b$ is also approximately zero. From this we calculate the internal torque (i.e the torque generated by muscle) at $x_b$:
\small
\begin{equation}
\begin{aligned}
&\vec{\tau}_{\mathrm{muscle}}+\int_{x_b}^{2\pi} \vec{r} \times \vec{f} \, \mathrm{d}s  = \dot{\vec{L}}\approx0 \\
&-\tau_{\mathrm{muscle}}=\tau(x_b,t) \\
&=\int_{x_b}^{2\pi}{\{(z-x_b)f_y(z,t)-[y(z,t)-y(x_b,t)]f_x(z,t)\} \, \sqrt{1+y_b'^2}\mathrm{d}z }.
\end{aligned}
\label{torque}\end{equation}
\normalsize \noindent where $\vec{f}$ is the granular force per unit length. We assume the muscle must only overcome torque from resistive forces $\tau$ and thus internal passive body forces are small compared to external resistive forces. This assumption was tested by performing \emph{in vivo} bending tests on an anaesthetized animal (see Methods and SI)and measuring stiffness and damping coefficients at varying rotation rates; we estimate that the maximal torques from elastic (0.094 $\pm$ 0.027\,N$\cdot$cm) and damping (0.055 $\pm$ 0.034\,N$\cdot$cm) forces are over an order of magnitude smaller than the maximal torque from resistive forces (4.1\,N$\cdot$cm). We also assume the time lag between neural activation and muscle force development is small compared to the sandfish undulation period ($\approx$ 0.5\,s). We thus assume activation timing approximately corresponds to ``muscle'' torque timing. Therefore, we use the sign of $\tau$ to predict muscle activation (Figs.~\ref{fig:model_diagram}\,\&\,\ref{fig:model}): positive $\tau$ (or negative $\tau_{\mathrm{muscle}}$) corresponds to muscle activation on the right side of the body.

\section{Results and Discussion}

\begin{figure}
\begin{center}
\includegraphics[width=0.5\textwidth]{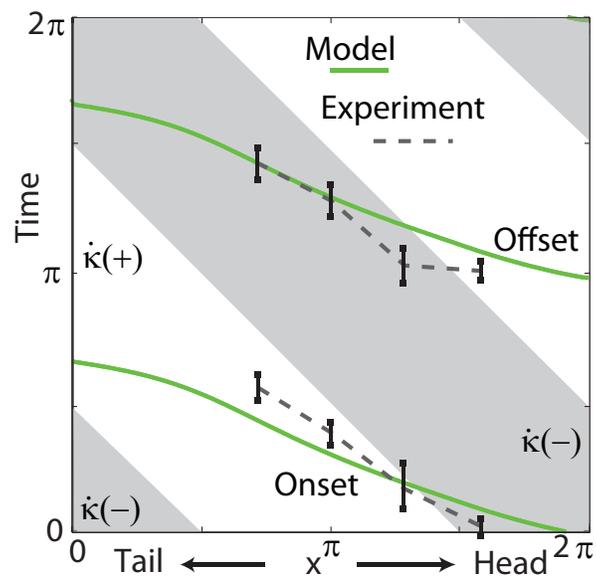}
\caption{The predicted onset and offset of muscle activation from the model (green lines) compared to EMG measurements from the sandfish experiment (black error bars indicates standard deviation, adapted from \cite{sharpe12}). Gray areas indicate the periods of negative $\dot\kappa$. $A/\lambda=0.22$ and the model sandfish body is oriented at a downward entry angle of 22 degrees relative to the horizontal. The corresponding positions of the electrodes in the model are approximated based on the curvature phases. \label{fig:compare}}
\end{center}
\end{figure}

We find that phase lags between internal torque and curvature in the model can explain the NPL between electromyogram (EMG) and curvature seen in experiments. $\tau$ displays a traveling wave pattern and positive $\tau$ occurs in a range close to that of measured EMG activation (Figs.~\ref{fig:model}\,\&\,\ref{fig:compare}). Without corrections from body passive forces or consideration of muscle physiology or body structure, the average phase difference between the beginning and ending of positive $\tau$ in the model compared to EMG onset and offset in experiments is less than 5\%, where $2\pi$ is the range of possible phase lags (Video S1 in SI). A large portion of the positive torque region overlaps with the region where the curvature decreases (negative $\dot{\kappa}$), but the positive $\tau$ region lags the negative $\dot{\kappa}$ region near the head and leads it near the tail. The agreement between experiment and theory is striking, particularly because our model has no fitting parameters; we posit this is largely a consequence of the simple movement and the relatively simple but strong environmental interaction.

To gain more insight into how the phase lags arise due to torque contributions from different parts of the body, we consider a simplified case where amplitude is small, forward motion is negligible, and the resistive force is viscous. This makes analytical calculation of torque straightforward but does not change the results qualitatively. In this simpler case, the torque from the fore-aft forces is negligible, and only the lateral force (per unit length) $f_y(x,t)=-c\dot{y}(x,t)=-cA\cos(x+t)$ need be considered.

For simplicity and to separate the effects, we first neglect yaw motion. The torque can be calculated analytically from Equation~\ref{torque}: $\tau(x_0,t)=(2 \pi-x_0) \sin(t)-\cos(x_0+t)+\cos(t)$. For example, if we take two points $x_1=\pi$ and $x_2=\pi-\Delta$ near the middle of the body, we obtain $\tau_{1}=3.7Ac\sin(t+\phi)$ and $\tau_{2}=(3.7+1.7\Delta)Ac\sin(t+\phi-0.29\Delta)$, where $\phi=0.57$. The NPL is still captured since the phase difference between $\tau_2$ and $\tau_1$ is a fraction (0.29) of $\Delta$, the phase difference between $\kappa_1$ and $\kappa_2$. The torque contribution can be approximately divided into three parts, as follows:

\begin{equation}
\begin{aligned}
\tau_{1}=&\int_{x_1}^{2\pi}{f_y(z,t)(z-x_1) \, \mathrm{d}z} \\
	\approx & \underbrace{\delta f_y(x_1,t)\delta /2}_{\mathrm{local}}+\int_{x_1+\delta}^{2\pi-\delta}{f_y(z,t)(z-x_1) \, \mathrm{d}z}\\
	& +\underbrace{\delta f_y(2\pi,t)(2\pi-x_1)}_{\mathrm{head}} \\
\tau_{2}=&\int_{x_2}^{2\pi}{f_y(z,t)(z-x_2) \, \mathrm{d}z}  \\
	\approx & \underbrace{\delta f_y(x_2,t)\delta/2}_{\mathrm{local}}+ \int_{x_2+\delta}^{2\pi-\delta}{f_y(z,t)(z-x_2) \, \mathrm{d}z} \\
& +\underbrace{\delta f_y(2\pi,t)(2\pi-x_2)}_{\mathrm{head}}  \\
\end{aligned}
\end{equation}

\begin{figure}
\centering
\includegraphics[width=0.8\textwidth]{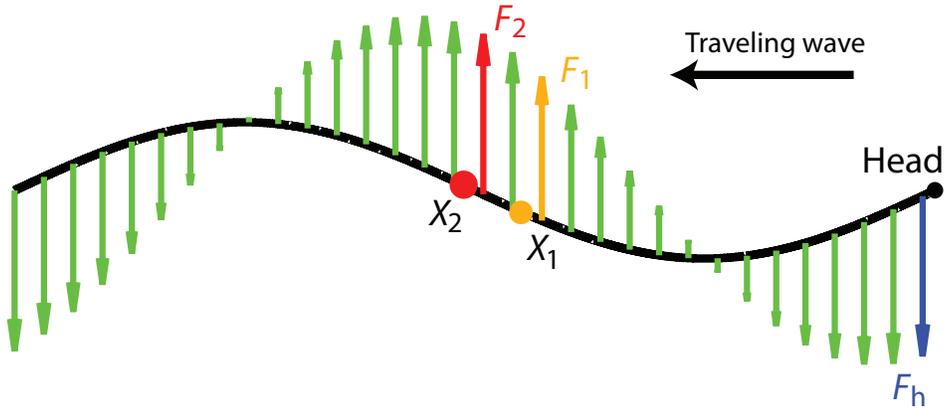}
\caption{Local and distant forces contribute to the torque on two points near the middle of the body. Green arrows represent the forces on the body. The red and orange arrows indicate the force adjacent to $x_2$ and $x_1$, respectively. The blue arrow indicates the force at the head ($F_h$). Note that this is an analysis for a small amplitude case, and the lateral displacement is exaggerated in the figure for visibility. \label{explanation}}
\end{figure}

\normalsize \noindent where $\delta$ is a small length. The phase difference between the torque contributions from local forces for the two points is $\Delta$, which is the same as the phase difference of other local variables (e.g. $\kappa$) on the traveling wave (Fig.~\ref{explanation}). In contrast, the phase of the torque transmitted from a distant point on the body (e.g. the head) is the same for both points (even though the magnitude differs). This synchronized torque contribution can be thought of as either a standing wave or a traveling wave with infinite speed. Because of the combination of the torques from local and distant forces and the continuous force distribution, the net phase difference between $\tau_2$ and $\tau_1$ is less than $\Delta$ and the torque wave speed is greater than the curvature wave speed. A similar analysis can be performed if the integration is done on the posterior side of the body (toward the tail).

\begin{figure}
\begin{center}
\includegraphics[width=0.95\textwidth]{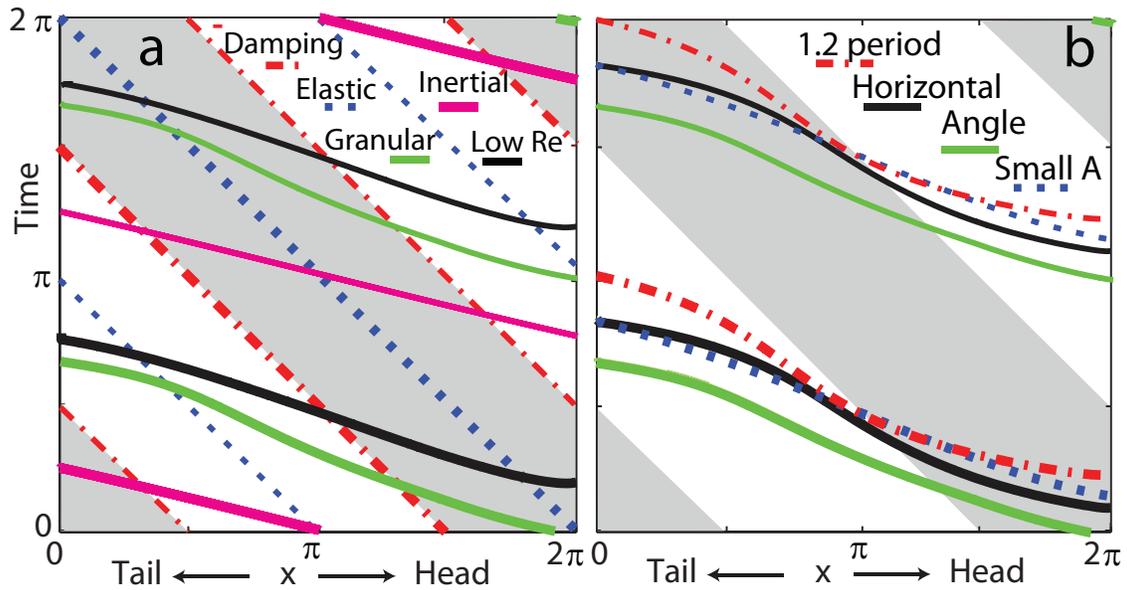}
\caption{NPL for varying model parameters. (a) The starting (slightly thicker lines) and ending of positive torque generated by granular force (solid green lines), viscous fluid force (solid black lines), inertial force (solid magenta lines), damping force (dash-dotted red lines), and elastic force from the body (dotted blue lines). (b) The beginning and ending of positive torque when the model sandfish swims in a horizontal plane (solid black line, control case), at an entry angle of 22 degrees (solid green line), at a small amplitude $A/\lambda=0.05$ (dotted blue line) , and with a 1.2 periods of wave on its body (dash-dotted red line). We aligned the head for the 1.2 period case and only the anterior portion (1 period of the wave) is shown in the figure. Gray areas indicate negative $\dot\kappa$. \label{fig:variation}}
\end{center}
\end{figure}

The balance of torque on the body leads to an overall yaw motion, whose phase is the same along the body. Superposition of yaw motion and lateral motion of the body results in variation of both the magnitude and phase of the lateral motion along the body in the lab frame (see Fig.~S2 and derivation in SI). However, the overall speed of the lateral displacement in the lab frame is the same as the prescribed lateral displacement (sinusoidal wave) in the body frame. Therefore, the yaw motion only changes the relative phase between the curvature wave and the apparent displacement (or force) wave locally.
Since the only requirement for this mechanism is a traveling wave pattern of force, it predicts the NPL are general for torques from distributed forces. As shown in Figure \ref{fig:variation}a, the localized elastic and damping forces by definition have constant phase differences with the curvature. In accord with previous studies~\cite{bowtell1991anguilliform,mcmillen2008nonlinear}, our calculations show that the relative phase between the torque from inertial forces and curvature advances in the posterior direction. However, the overall phase of the inertial torque advances by about $0.4\pi$ compared to the sandfish EMG phase. The phase lags persist if the granular resistive forces in the model are replaced with viscous resistive forces, which low Reynolds number swimmers like nematodes experience~\cite{cohen2010swimming}.

Although passive body forces are not responsible for the NPL, they can still influence the observed pattern. For example, we find that the inclusion of viscous forces in the body shifts the phase of the torque in granular media toward the phase pattern produced from only viscous forces (dash-dotted red lines in Fig.~\ref{fig:variation}a). That is, the phase difference between the torque and $\dot{\kappa}$ is smaller and the torque wave speed is smaller, in accord with previous studies in fluids~\cite{cheng1998continuous,tytell2010interactions}. This suggests that the small internal viscous forces within the body may partially account for the phase differences we observe between the torque from resistive forces and EMG.  For swimming in a high Reynolds number fluid, the muscle activation duration is in general smaller than those observed for the sandfish ($\approx 0.5$)~\cite{wardle1995tuning}. Previous studies (e.g. \cite{cheng1998continuous}) suggest that the torque from external forces may be overcome by passive elements of the body. The nearly 0.5 duty factor of the muscle is evidence that resistive forces dominate in a granular environment and the slight decrease of the duty factor (a relatively larger decrease is typical during swimming in fluids~\cite{wardle1995tuning}) implies passive forces play a small role for swimming of the sandfish.

Variations of locomotor kinematics also affect the timing of the torque (Fig.~\ref{fig:variation}b). For example, a downward entry angle (observed in the animal experiments~\cite{sharpe12}) advances the phase of the torque compared to the horizontal swimming case. This is because when the body is oriented downward, the head, which has a more advanced phase, contributes more to the overall torque due to the head's greater depth and correspondingly larger resistive forces.  Also, a larger number of periods (longer body and smaller ratio of wave length to body length) both delays the phase of torque and reduces the torque wave speed.  The phase shift is due to the contribution of the extra tail length, where the phase of the force lags that at anterior positions. The effect of period (body length) can be used to estimate the error in timing that may occur due to neglecting the tail after 1.2 SVL: The error should be a small fraction of the difference between the 1.2-period case and the control case. Further, we found that a smaller undulation amplitude reduced the variation in torque wave speed.

The time delay between EMG  activation and force production (\cite{daley2003}) might affect the phase lag timing of EMG  activation, but we argue that this delay is small compared to the typical period of undulation for the sandfish ($\approx$2\,ms latency compared to $\approx$500\,ms undulation period). If the time delay was significant and approximately constant, the EMG-curvature phase relation would change for different frequencies.

\section{Conclusions}

We developed a theory to explain the basic control signals needed to generate a particular undulatory movement pattern in a sand-swimming lizard. We abstracted the nervous/musculoskeletal system by assuming that passive body forces are small and that internal torque is synchronized with neural activation timing; this abstraction revealed that the NPL is intrinsic to undulatory locomotion provided that distributed forces, such as resistive or inertial forces, play major roles. For undulatory locomotion in other environments, the principle of the simultaneous response to distant torques should also apply, though quantities such as the phase of the force will differ from the sandfish case. Building on this principle could help future studies explain other variations of the NPL.

Because we now have a system in which experiment and theory are in quantitative agreement, we can begin to develop more detailed models (i.e. anchors~\cite{fulAkod})which answer specific questions about nervous system control, muscle configuration, morphology, etc.  For example, it has been established that the intersegmental coordination of neural oscillators along the body of swimmers is influenced by sensory feedback (\cite{marder2005}). Detailed models of central pattern generators (CPGs), sensory neurons and muscles can be used to understand how external torque and neural activation interact so that the intersegmental phase lags produce \emph{single period} sinusoidal motion.  As such, a hierarchy of anchors can be used to generate testable hypotheses and understand actuation timing for animals in a variety of environments.

More broadly, we have demonstrated that the seemingly specific and peculiar sand-swimming behavior could be an excellent system in which to develop quantitative models of neuromechanics. Due to relatively simple but dominant environmental interactions, the neuromechanical control pattern is greatly constrained by the environment. In addition, the granular RFT provides an excellent model for interaction with the substrate; this is in contrast to locomotion in true fluids in which more complex theories~\cite{Rodenborn2013propulsion} are needed to quantitatively compare experiment and model. We hypothesize that by studying subarenaceous animals within dry and saturated granular substrates (like those on the bottom of the ocean floor), animal models with potentially fewer parameters can be analyzed in detail. This in turn can help provide guidance for the design and control of artificial undulatory locomotors in complex environments~\cite{hirosebiologically,roper2011review,colgate2004mechanics}. Better physical models can also improve our understanding of the biological systems.

\section{Materials and Methods}
\subsection{EMG Recordings}
Previous work \cite{sharpe12} using a micro-CT scan of a single sandfish revealed 26 vertebrae in the trunk and more than 13 anterior caudal vertebrae in the tail.  The iliocostalis musculature was targeted for implantation and is located on the dorso-lateral portion of the trunk. Dissection revealed qualitatively similar muscle morphology to that described for \textit{Iguana iguana}~\cite{Carrier1990,Ritter1996}; where iliocostalis musculature spanned approximately 1 vertebrae.

Electrodes were implanted in one side of the body at 0.3 (magenta), 0.5 (green), 0.7 (blue), 0.9 (yellow) snout-vent length (SVL)(Fig.~\ref{fig:intro}b) where the average SVL was 8.9\,cm (N = 5 animals). EMG data used in this paper were taken from n=37 sandfish swimming trials. The EMG signal was filtered with a second-order Chebyshev filter and rectified in order to facilitate EMG burst detection. A burst threshold was set equal to the mean of this rectified-filtered EMG trace. Burst onset was defined as the time when the filtered EMG signal exceeded the threshold and afterwards remained above it for a minimum of 0.04s. EMG burst offset was defined as when the filtered EMG signal became lower than the threshold and remained below for at least 0.08\,s ~\cite{hochman2012enabling}.  This burst detection was necessary to exclude small voltage changes that did not constitute an EMG burst, such as noise due to movement artifact. See \cite{sharpe12} for more details on the EMG recording and analysis technique.

\subsection{Dynamic Bending Tests} 
\begin{figure}[htp]
\centering
\includegraphics[width=0.60\textwidth]{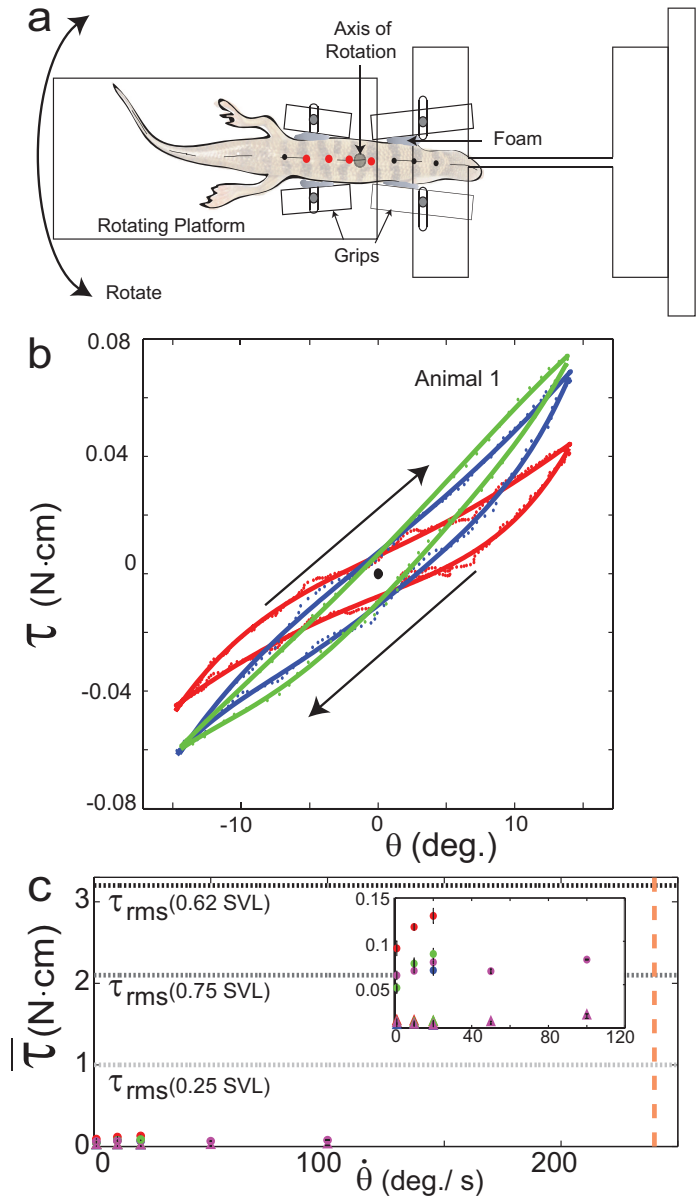}
\caption{Experimental measurements of sandfish body elasticity and damping. (a) Top view of the setup. (b) Representative work loop for Animal 1 with angular velocities of $1^\circ/\mathrm{s}$ (red), $10^\circ/\mathrm{s}$ (blue) and $20^\circ/\mathrm{s}$ (green). The direction is indicated by the arrows, where the angle is initially zero and increases to $\approx 0.26$ rad. ($15^\circ)$ , decreases to $\approx -0.26$ rad. ($-15^\circ)$, then returns to zero. Work loops are shown for constant angular velocity (i.e. the system has reached a steady state). Dotted trajectories represent the experimentally recorded force; solid curves are the best polynomial fits. (c) Estimated average torque using calculated elastic (circles) and damping (triangles) coefficients compared to the maximum $\tau_{\mathrm{rms}}$ due to external forces calculated from discrete element simulation ~\cite{ding2012mechanics} (0.62 body length (BL))(dashed horizontal black line).  Average $\tau_{\mathrm{rms}}$ for 0.25 BL(light gray dashed line) and 0.75 BL (dark gray dashed line) are also shown. For animal 1 (red), 2 (blue) and 3 (green) coefficients were measured for 1, 10 and 20$^\circ/\mathrm{s}$.  In animal 4 (magenta) coefficients were measured for 1, 10, 20, 50 and 100 $^\circ/\mathrm{s}$. The average angular speed the sandfish operates at is 240 $^\circ/\mathrm{s}$. Inset shows zoomed in region of data in the main figure (units are the same). \label{fig:bending}}
\end{figure}

Three anesthetized sandfish (mass = 15, 16 and 25\,g) were gently clamped at approximately 0.5 SVL and at 0.6 SVL (Fig.~\ref{fig:bending}a) with adjustable grips. The grips were attached to a rigid platform and to a rotating platform, respectively. A motor rotated the anterior region of the sandfish through $\pm$ 15$^\circ$ for 3 cycles at angular velocities of 1, 10 and 20$^\circ$/s. The first and last half cycle were excluded from the analysis due to varying rotation velocities. The anterior end of the sandfish was clamped to a platform with two strain gages (Omega, KFG-3-120-C1-11L1M2R) used to record resulting torques. Signals were amplified (INA125P; Digi-Key) by 5000 before data acquisition and analyzed using custom software (LabVIEW, NI, Austin, TX, USA). Black points were marked on the animal midline at increments of 0.1 SVL. The best fit line through the markers circled in red were used to calculate the angle $\theta$.

Body stiffness, $K$, was estimated using the slope of the best fit line through the torque-angular displacement curve (Fig.~\ref{fig:bending}b) for a single cycle (n=8 trials each). Using a viscoelastic model (or Voigt model), the viscous damping coefficient, $c$, was approximated by quantifying the viscous torque ($\tau_v$) at zero angular displacement during steady-state rotation and dividing by the angular speed ($\dot{\theta}$), (i.e. $c = \tau_v(\theta=0)/{\dot{\theta}}$).

For the hysteretic damping model, the structural damping coefficient $h$ was proportional to angular displacement, $\theta$ and $\pi/2$ out phase. The loss factor $\eta = h/ K$. $h$ was estimated by finding the torque at zero displacement during steady state rotation and dividing by the maximum angular displacement:  $ h = \tau_{h}(\theta=0) /{\theta}_{\mathrm{max}}$. The area contained within the work loop ($E_{\mathrm{loss}}$) was determined using polynomial fits to the torque vs angle curves for increasing and decreasing angle.

To interrogate stiffness and damping coefficients at higher speeds, we repeated the experiment with one of the sandfish (animal 2, Fig. S3) using angular speeds of 1, 10, 20, 50 and $100^\circ$/s and compared with previous results.

We substituted the values during sand-swimming (angular excursion of 30$^\circ$ and angular velocity of 240$^\circ$/s) and the average calculated $K$ and $c$ at 20$^\circ$/s into our viscous damping model to estimate torques during sand-swimming (Figs.~\ref{fig:bending}c\,\&\,S4). For hysteretic damping, we estimated the damping torque at $240^\circ$/s by extending the trend line between calculated torque from hysteretic damping and angular speed between 20 and 100$^\circ$/s.

\subsection{Pendulum swing tests} In the second technique, sandfish were modeled as a physical pendulum to estimate $K$, $c$ and $\eta$ at higher angular frequencies. The same sandfish were used as in the previous experiment.  Animals were oriented vertically and clamped at approximately 0.5 SVL. The tail of the sandfish was bent upward and released, allowing the body to swing freely. The sandfish body was modeled as a rigid cylinder and the tail as a cone with uniform density. The angular motion, $\theta$ was fit to a damped harmonic oscillator:

\begin{equation}
I\ddot{\theta}+ c\dot{\theta}+K\theta+mgd_{\mathrm{CoM}}\sin(\theta)=0,
\label{sine}\end{equation}

\noindent where $d_{\mathrm{CoM}}$ is the distance from the point of rotation to the center of mass, $m$ is the mass of the unclamped portion of the sandfish, $I$ is the moment of inertia, and $\ddot{\theta}$ is the angular acceleration. Angular motion during the first half cycle after the tail was released was neglected due to large angles and body bending. $\theta$ was measured between the 0.5 and 0.8 SVL body positions.

We also fit the motion using a hysteretic damping model:
\begin{equation}
I\ddot{\theta}+ (1+ i\eta)K\theta+mgd_{\mathrm{CoM}}\sin(\theta)=0.
\end{equation}

\noindent For both models, we used the small angle approximation $\sin(\theta)\approx \theta$. Best fit parameters were determined using minimization techniques (Matlab, Mathworks, Natick, MA, USA). Both viscous and hysteretic models fit the angular displacement trajectory well ($r^2 < 0.9$).  See SI for experimental setup diagrams and detailed results.

\section{Acknowledgments}
We thank Paul B. Umbanhowar, Silas Alben, George Lauder, Tom Daniel, and Robert J. Full for helpful discussion and Humaira Taz for assistance with construction of experimental apparatus. We thank Elizabeth A. Gozal for providing the EMG analysis code (SpinalMOD). This work was supported by NSF Physics of Living Systems (PoLS) grants No. PHY-0749991 and No. PHY-1150760, the ARL MAST CTA W911NF-11-1-0514, and the Burroughs Wellcome Fund.

\bibliographystyle{pnas}
\bibliography{masterbib}
\end{document}